# Reconfigurable Curved Beams at Terahertz Frequencies Using Inverse-Designed Bilayer Diffractive Structures

Wei Jia*, Miguel Gomez, Steve Blair, and Berardi Sensale-Rodriguez*
Department of Electrical and Computer Engineering, The University of Utah, Salt Lake City, UT, 84112, USA
Corresponding authors: happywei.jia@utah.edu, berardi.sensale@utah.edu

**Abstract**: Curved electromagnetic beams at terahertz (THz) frequencies have recently emerged as a powerful example of wavefront engineering, with applications in imaging and high-capacity wireless communications. Unlike canonical self-accelerating solutions such as Airy beams, general curved-beam propagation enables arbitrary, application-specific trajectories that are not constrained by analytic beam families. Here, we demonstrate a passive and reconfigurable approach for generating trajectory-engineered THz curved beams using inverse-designed bilayer diffractive optical elements (DOEs). Two phase-only diffractive layers are optimized using gradient-based inverse design to produce predetermined curved propagation paths. Reconfiguration is achieved by a 180° rotation of the second layer, which modifies the effective phase profile of the cascaded structure without altering the incident wave or individual layer designs. The proposed system can produce distinct curved trajectories with controlled transverse displacement and beam confinement, as confirmed by scalar diffraction simulations and experimental measurements. Overall, this work establishes inverse-designed cascaded DOEs as a compact and scalable platform for reconfigurable trajectory control of THz beams, providing a flexible alternative to analytic self-accelerating beams for radiative near-field THz communications.

1. Introduction

The terahertz (THz) frequency range (0.1 ~ 10 THz) has been attracting significant attention in both research and industrial applications due to its unique potential for non-destructive testing [1,2], biomedical imaging [3,4], security screening [5], and high-capacity wireless communication [6,7]. Many of these applications require dynamic control of the THz wavefronts. However, implementing reconfigurable THz optics remains challenging, partially due to the need for active materials, which often require complex actuation.

More broadly, THz systems are entering a regime in which controlling how energy flows through space, along engineered trajectories, has become a powerful degree of freedom [8].

Curved-beam propagation in the radiative near-field offers a fundamentally different approach beyond conventional beam steering, enabling beams to reach locations that are inaccessible to straight-line paths without relying on scattering or reflections. This shift from angular steering to trajectory engineering motivates new optical architectures capable of synthesizing programmable application-specific propagation paths.

Diffractive optical elements (DOEs) offer an efficient way for electromagnetic wavefront engineering across a wide range of frequencies, including the THz regime [9–11]. By locally controlling phase or amplitude, the diffractive structures can achieve compact and lightweight beam shaping functionalities [12]. To achieve reconfiguration, passive approaches through spatial transformations, such as translation or rotation, offer an attractive alternative due to their simplicity [13].

Recent work has highlighted the importance of curved and self-accelerating beams in the THz regime, particularly for radiative near-field wireless communication applications where line-of-sight blockage and dynamic environments limit conventional beam steering [14]. In this context, curved THz beams have been experimentally shown to route data-carrying wavefronts around obstacles by exploiting near-field interference rather than ray-optical steering [14–16]. While such demonstrations mainly rely on Airy beams as analytically defined self-accelerating solutions, such beams inherently follow parabolic trajectories dictated by a small set of parameters, which constrain flexibility in practical environments. More broadly, wavefront engineering in the radiative near-field enables an effectively unbounded space of possible beam trajectories, provided that the aperture phase can be tailored with sufficient precision. This motivates the development of inverse-designed diffractive optics capable of synthesizing arbitrary curved propagation paths without relying on a specific analytic beam family. Here we adopt this general trajectory-engineering framework and demonstrate a passive, reconfigurable diffractive platform that produces multiple predetermined curved THz beams through spatial multiplexing of bilayer phase-only structures, extending curved-beam concepts beyond Airy-based implementations toward fully programmable near-field propagation.

Our work introduces a reconfigurable terahertz curved-beam platform based on spatial multiplexing of inverse-designed bilayer diffractive optical elements. Unlike Airy beams, which represent a specific self-accelerating solution to the paraxial wave equation, the proposed approach enables the synthesis of arbitrary curved trajectories defined directly in real space. The two diffractive layers are jointly optimized to produce distinct propagation paths, and reconfiguration is achieved through a simple 180° mechanical rotation of the second layer, effectively altering the composite phase response as schematically illustrated in Fig. 1. This strategy avoids the need for active materials or dynamic biasing while providing additional degrees of freedom for near-field wavefront control. The demonstrated capability to reconfigure predetermined curved trajectories highlights the potential of inverse-designed diffractive optical networks as a flexible platform for trajectory-aware THz imaging and communication systems.

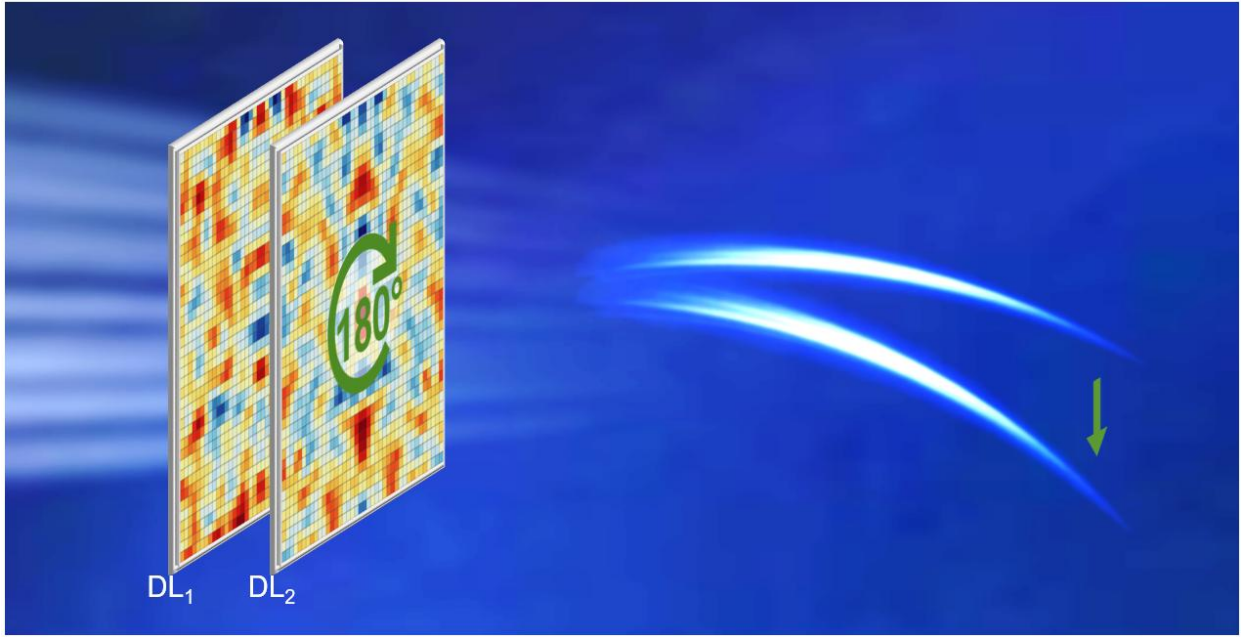

Fig. 1 Concept and operating principle of the reconfigurable trajectory-engineered terahertz curved-beam platform. A bilayer diffractive structure composed of two inverse-designed, phase-only diffractive layers ($DL_1$ and $DL_2$) shapes an incident collimated THz beam into a curved propagation trajectory in the near-field. When the two layers are aligned in their original orientation, the cascaded phase profile produces one predetermined curved beam. Rotating the second layer (DL2) by 180° modifies the effective composite phase distribution without changing the individual layer designs, resulting in a distinct curved trajectory. This passive spatial reconfiguration enables multiple trajectory-engineered propagation paths from the same diffractive platform.

## 2. Design and simulation

The proposed structure consists of two diffractive layers (DLs), denoted as $DL_1$, $DL_2$, each with a dimension of 80 mm × 80 mm. Both layers are discretized into 800 × 800 pixels, resulting in a pixel size of 0.1 mm × 0.1 mm. At the design frequency of 0.3 THz, corresponding to a free-space wavelength of approximately 1 mm, the chosen pixel size is deeply subwavelength, enabling effective wavefront control through phase modulation.

Rather than prescribing an analytic phase profile, the diffractive layers are designed by directly optimizing the desired field evolution along the target trajectories. That is, the two cascaded DLs are inverse designed with the goal of imparting spatially varying phase profiles that enable a controlled curved beam trajectory. When the two DLs are aligned in their original orientation, the structure produces one curved trajectory, as indicated by the blue curve in Fig. 2(a). Rotating DL$_2$ by 180° along the z-axis generates a different combined phase distribution, resulting in a distinct trajectory, shown by the red curve in Fig. 2(a). The spacing $d$ between the two layers is 20 mm and the two target curved trajectories are confined to the propagation range from $z_1$ = 45 mm to $z_2$ = 155 mm, with their paths defined in mm units by $y_1 = -2.69 \times 10^{-6} \cdot (z - 45)^3 + 3$ and $y_2 = -1.34 \times 10^{-5} \cdot (z - 45)^3 - 3$, as illustrated in Fig. 2(b). The corresponding transverse bending ranges along the y-axis are 3.6 mm and 17.8 mm, respectively. Along these trajectories, the beams are designed to maintain a "Gaussian"-like intensity profile with a beam waist of 1.5 mm, as shown in the insert of Fig. 2(b).

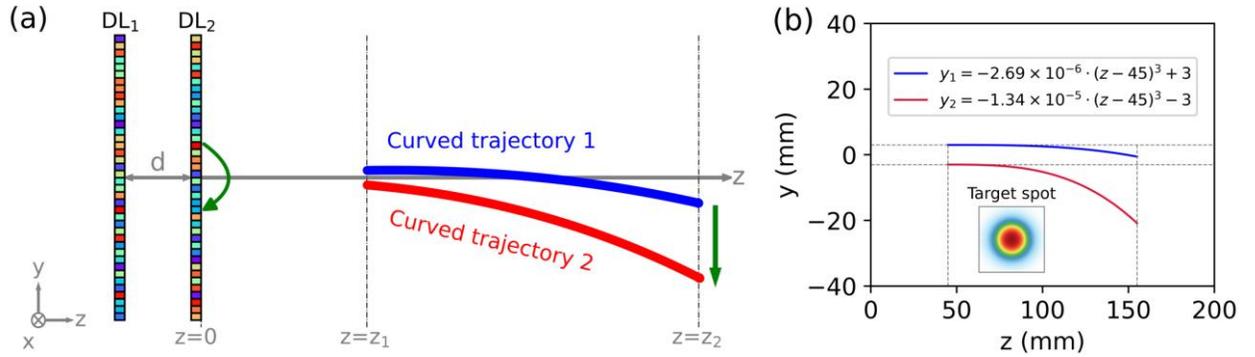

Fig. 2 (a) Schematic of reconfigurable curved beams enabled by spatial multiplexing of bilayer diffractive structure. (b) Curved beam trajectories, with an insert indicating the target spot along the paths.

PyTorch is employed to optimize the phase distributions $\phi_1(x, y)$ and $\phi_2(x, y)$ on the two DLs. The corresponding complex transmission coefficients are denoted by $t_1(x, y)$ and $t_2(x, y)$, which are related to the phase profiles through $t(x, y) = \exp[j\phi(x, y)]$. Wave propagation is modeled using scalar angular spectrum diffraction theory [17], given by:

$$E(x, y, z) = \mathcal{F}^{-1}\{\mathcal{F}[t(x, y)] \cdot H(f_x, f_y, z)\},$$

where $\mathcal{F}$ and $\mathcal{F}^{-1}$ denote the fast Fourier transform and its inverse, respectively. The angular spectrum transfer function is

$$H(f_x, f_y, z) = \exp\left(j\frac{2\pi}{\lambda} \cdot z \cdot \sqrt{1 - (\lambda f_x)^2 - (\lambda f_y)^2}\right),$$

with $\lambda$ being the wavelength, $z$ the propagation distance, and $f_x$ and $f_y$ the spatial frequency coordinates. The loss function used for optimization is defined as:

$$L = \frac{1}{m}\sum_{i=1}^{m} \frac{w_{1,i} \cdot MSE(I_{1,z_i}, T_{1,z_i}) + w_{2,i} \cdot MSE(I_{2,z_i}, T_{2,z_i})}{2}$$

where $w_{1,i}$ and $w_{2,i}$ are weighting coefficients. Here, $I_{z_i} = |E_{z_i}|^2$ is the diffracted intensity distribution at the target plane located at $z_i$. The target planes are uniformly sampled over the range $z_i \in [45, 155]$ mm with a step size of 5 mm, resulting in $m = 23$ planes. The operator $MSE(\cdot)$ denotes the mean squared error between the simulated intensity $I_{z_i}$ and the target intensity distribution $T_{z_i}$.

The phase distributions $\phi_1(x, y)$ and $\phi_2(x, y)$ are parameterized as learnable tensors in PyTorch, enabling efficient gradient based optimization. Starting from an initial guess, the phases on two DLs are iteratively updated via backpropagation to minimize the defined loss function $L$. In each iteration, an incident Gaussian beam with a beam waist of 30 mm is propagated through both DLs. The intensity distributions at each target plane are computed using the angular spectrum method for two configurations of the second DL rotation along z-axis: 0° and 180°. The resulting intensities for curved trajectories are compared to their corresponding target patterns by evaluating the defined loss function. PyTorch's automatic differentiation engine, coupled with the Adam optimizer, calculates gradients of the loss with respect to the phase parameters, which are then used to update the phase profiles of the DLs. During the optimization, symmetry is imposed along the y axis, reducing the number of independent learning parameters to 800 × 400 for each DL. Each layer is then zero-padded by 400 pixels on all sides to expand the computational window to 1600 ×1600, thereby reducing artificial boundary interactions caused by the fast Fourier transform and enable more accurate modeling of the propagation of the scalar diffraction beam. The optimization proceeds for a maximum of 1000 iterations and completes in around four minutes with the acceleration of GPU NVIDIA RTX A2000. Finally, the optimized phase distributions are converted into height profiles according to

$$h(x, y) = \frac{\phi(x,y)\lambda}{2\pi[Re(n)-1]},$$

where $n = 1.68 + j0.03$ is the refractive index of the resin used for 3D printing (ABS-like photopolymer resin) and Re(n) indicates its real part. The resulting height distributions of the two DLs are shown in Fig. 5(a). To enable robust fabrication, a uniform base thickness of 1 mm is added to each diffractive layer. All subsequent wave-propagation simulations for intensity distributions are performed using these physical height profiles, including the added base thickness, thereby accounting for material absorption losses.

The simulated intensity distributions along the propagation axis (z) for two distinct curved beams are shown in Figs. 3 and 4. Figure 3 depicts curved beam 1, with its volumetric 3D intensity distribution in Fig. 3(a) illustrating a controlled, moderately curved trajectory resulting from the optimized phase profiles. The corresponding y–z cross-section in Fig. 3(b) and the transverse intensity profiles at multiple propagation distances in Fig. 3(c) confirm the beam closely follows the designed path with the intended shape and intensity distribution.

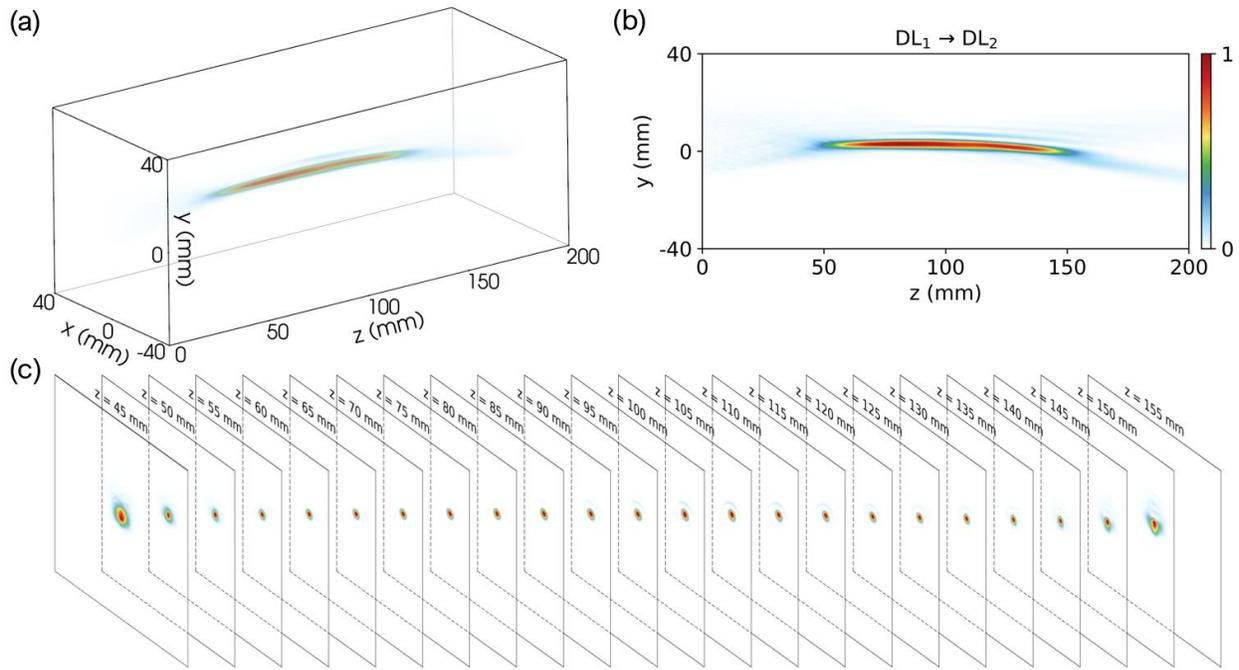

Fig. 3. Simulated intensity distribution of curved beam 1: (a) volumetric view, (b) y–z plane, (c) x–y plane at different propagation distances z.

In contrast, Fig. 4 shows curved beam 2, which exhibits a more pronounced curvature due to the second DL rotated by 180° along z-axis. The volumetric view in Fig. 4(a) reveals this stronger curved trajectory, further confirmed by the *y–z* slice in Fig. 4(b). The transverse intensity profiles across various propagation distances in Fig. 4(c) demonstrate that beam 2 maintains its desired intensity distribution and shape throughout its more curved path. These results collectively validate the ability of the optimized phase profiles to generate distinct and reconfigurable curved behaviors with spatial multiplexing of the two diffractive layers.

At this point, it is important to highlight that although this work demonstrates two distinct curved trajectories enabled by a bilayer diffractive structure, the inverse-design framework is not limited to two layers or two configurations. Additional diffractive layers or alternative spatial multiplexing strategies could be incorporated to increase the number of accessible beam trajectories, enabling multi-state or continuously reconfigurable propagation control. The primary constraints on scalability are computational complexity and fabrication tolerances, rather than fundamental physical limitations of the approach.

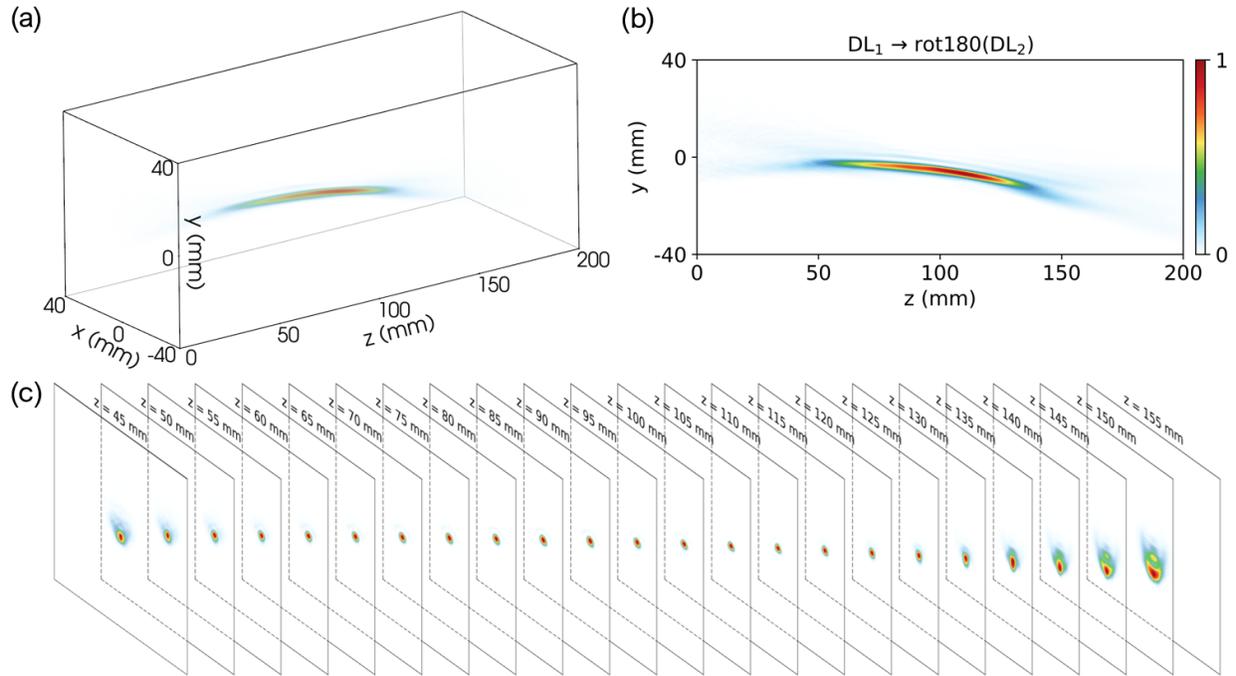

Fig. 4. Simulated intensity distribution of curved beam 2: (a) volumetric view, (b) y–z plane, (c) x–y plane at different propagation distances z.

## 3. Fabrication and measurement

A Saturn 4 Ultra 16K 3D printer was utilized to fabricate the designed optical component through 3D printing technique with clear color water washable resin. During the printing, the diffractive layers are aligned vertically to the resin tank. The top-view of the fabricated two diffractive layers is shown in Fig. 5(b).

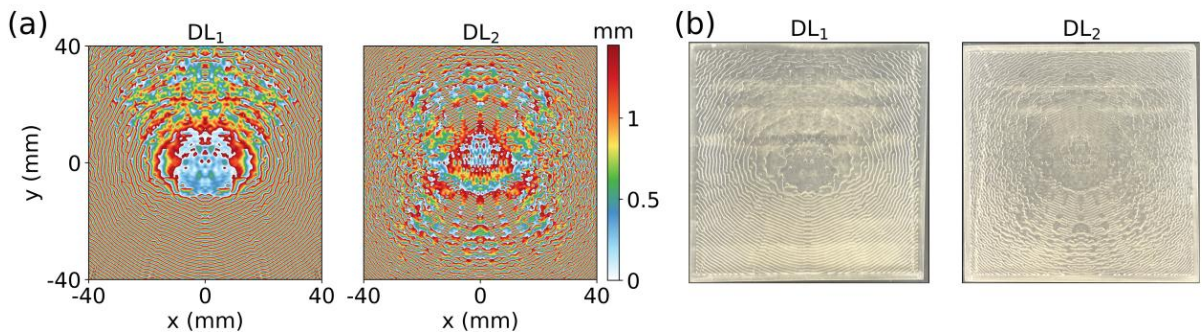

Fig. 5(a) Optimized height distributions of the two diffractive layers. (b) 3D printed diffractive layers with clear color water washable resin.

The experimental characterization setup is shown in Fig. 6(a). A 0.3 THz wave is emitted by the transmitter through a horn antenna and subsequently propagates through the first Fresnel lens (FL$_1$), after which the beam diverges toward the second Fresnel lens (FL$_2$). FL$_1$ has a diameter of 75 mm and a focal length of −40 mm, while FL$_2$ has a diameter of 150 mm and a focal length of 90 mm and is used to collimate the beam. The simulated spatial intensity distributions of FL$_1$ and FL$_2$ in the y–z plane are shown in Fig. 6(b). Figure 6(c) shows the two lenses fabricated using Original Prusa MK4S 3D Printer with high-impact polystyrene (HIPS) material.

The collimated beam emerging from FL$_2$ serves as the incident field for the fabricated bilayer diffractive structure, consisting of DL$_1$ and DL$_2$. Transmission through this diffractive structure produces a focused and curved beam trajectory. The spatial intensity profile of the resulting beam was experimentally measured over a focal range extending from $z_1$ = 45 mm to $z_2$ = 155 mm. Intensity measurements were performed using a THz Tera 1024 camera, which incorporates a 32 × 32 pixel sensor array with a pixel pitch of 1.5 mm. The camera was mounted on a precision three-axis translation stage to enable further spatial alignment and positioning.

To facilitate systematic scanning along the beam propagation direction, a motorized linear translation stage was used to automatically translate the camera across the measurement range. This arrangement enables precise and repeatable acquisition of the evolving intensity distribution of the curved beam throughout the focal region.

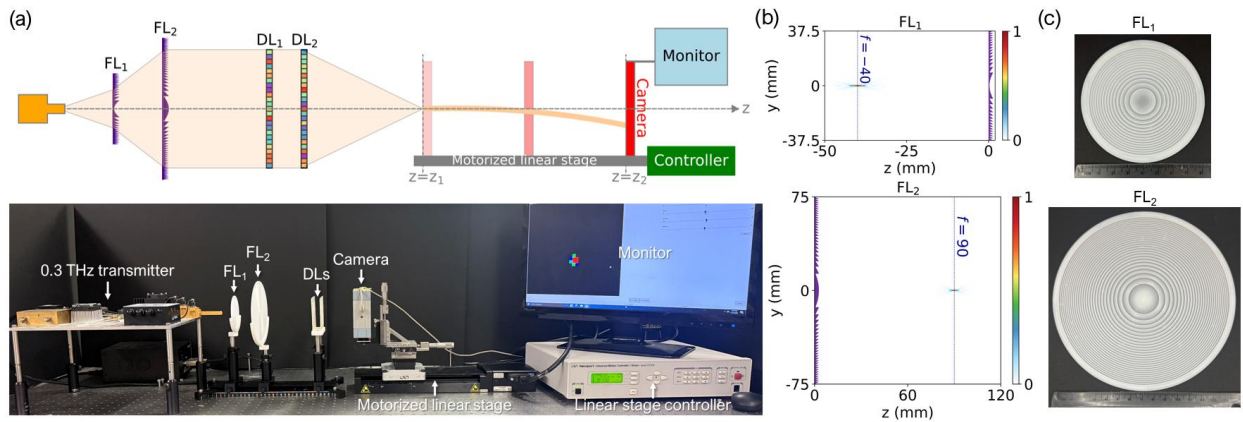

Fig. 6 (a) Schematic diagram and photograph of the experimental measurement setup. (b) Simulated intensity distribution in the y–z plane for Fresnel lenses FL$_1$ and FL$_2$. (c) 3D printed Fresnel lenses of FL$_1$ and FL$_2$ with HIPS material.

## 4. Results and discussion

The ability to generate and reconfigure predetermined curved terahertz beam trajectories has implications beyond static beam shaping. In wireless systems, such control enables trajectory-aware links that can route energy around obstacles or dynamically adapt to changing environments. In imaging and sensing, curved beams can probe regions that are shadowed or inaccessible to conventional Gaussian illumination, potentially improving coverage and robustness. Importantly, the passive and planar nature of the proposed diffractive platform makes it compatible with large-aperture implementations, where diffraction-dominated effects are most pronounced. These capabilities position inverse-designed diffractive optics as a key enabler for trajectory-controlled THz systems.

In what follows, we present experimental results that validate this trajectory-engineering concept. We first compare the measured beam evolution for the two reconfigurable states of the bilayer structure, corresponding to the two designed curved trajectories. We then analyze conversion efficiency and sensitivity to alignment tolerances, providing insight into practical considerations relevant for scalable implementations.

To experimentally probe the two reconfigurable propagation states, the second diffractive layer ($DL_2$) is first aligned with the first layer ($DL_1$) without rotation, the measured x-y plane intensity distributions of curved beam 1 at propagation distance ranging from 45 to 155 mm, in 5 mm increments, are shown in Fig. 7(a). These measurements were acquired using the THz Tera-1024 camera while translating the camera along the z direction with a motorized linear stage. The recorded measurement video is shown in SI1. When $DL_2$ is rotated by 180° along the z-axis and realigned with $DL_1$, the corresponding x-y plane intensity distributions of curved beam 2 at the same propagation distances are shown in Fig. 7(b) with the related measurement video shown in SI2.

To more clearly visualize the beam curved trajectories, the x-y plane intensity distributions at each propagation distance were averaged along the x axis. This averaging procedure suppresses local intensity fluctuations and detector noise, thereby providing a more robust representation of the overall beam envelope. The resulting y-z plane intensity distributions are presented in Fig. 7(c) and 7(d) for curved beam 1 and curved beam 2, respectively. From these reconstructed y-z profiles, a distinct difference in behavior is observed in that curved beam 2 exhibits a noticeably greater curvature than curved beam 1 over the same propagation distances. This enhanced curvature arises from the 180° rotation of $DL_2$, which alters the effective phase gradient imposed on the wavefront. Importantly, this experimentally observed trend is consistent with the designed goal of the bilayer diffractive structure, confirming that reconfigurable curved beam trajectories can be achieved through spatial multiplexing of a bilayer diffractive structure.

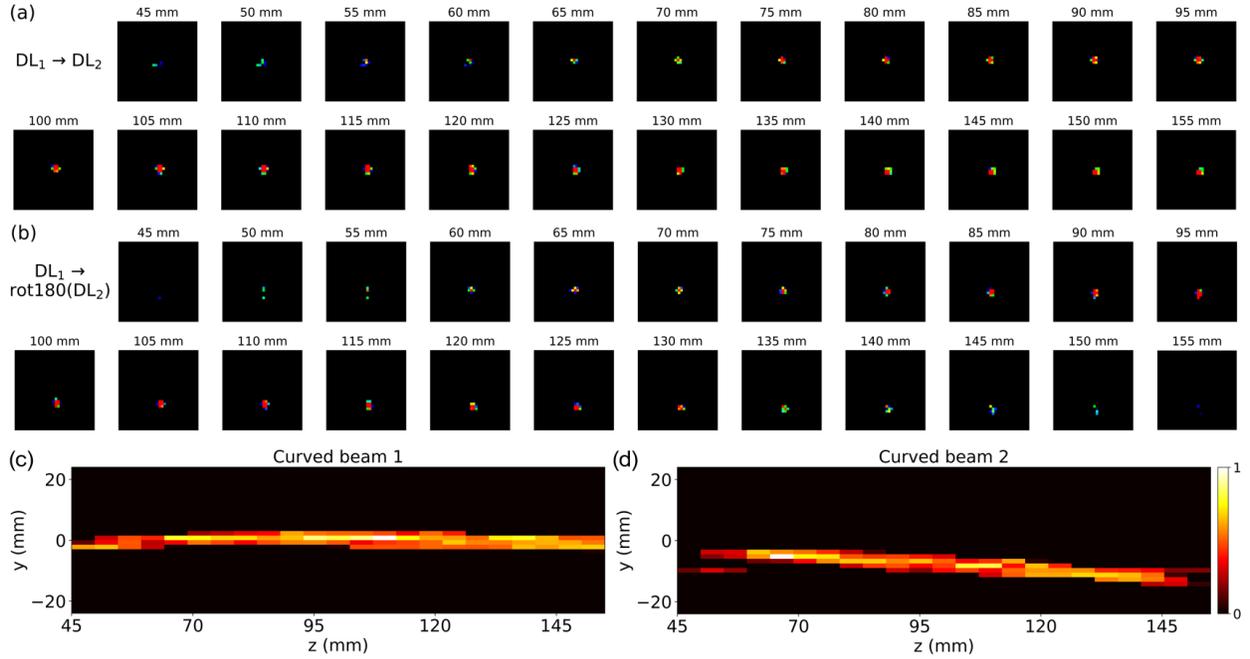

Fig. 7 (a) Measured x–y plane intensity distributions of curved beam 1 at propagation distances from 45 to 155 mm with a 5 mm step size. (b) Measured x–y plane intensity distributions of curved beam 2 over the same distance range. (c) y–z plane intensity distribution of curved beam 1 reconstructed by averaging the x–y data along the x direction. (d) y–z plane intensity distribution of curved beam 2 reconstructed using the same averaging method.

The simulated conversion efficiencies of the two curved beams are evaluated over a propagation range from 45 mm to 155 mm, as shown in Fig. 8(a). The conversion efficiency is defined as the ratio of the optical power contained within 1.5 times the target beam waist to the total incident power. When considering the resin material with a refractive index of $n = 1.68 + j0.03$, which includes both the 1 mm base thickness for each layer and the intrinsic absorption losses of the resin, the average conversion efficiencies for the two beams are around 10.6% and 10.5%, respectively. In contrast, assuming a lossless resin material with $n = 1.68 + j0.00$, the conversion efficiencies improve significantly, as illustrated by the dashed lines in Fig. 8(a). Under this ideal lossless resin material condition, the average efficiencies in the range of 45 to 155 mm rise to 38.8% and 37.8% for the two curved beams, respectively.

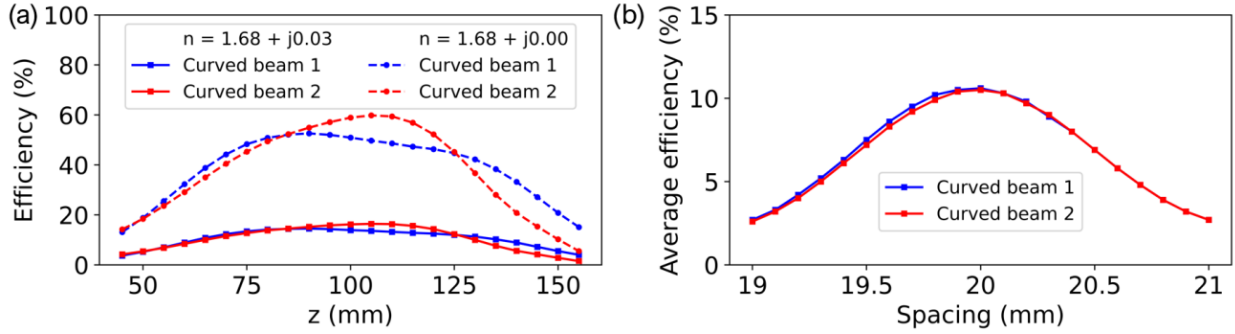

Fig. 8 (a) Conversion efficiencies of the two curved beams over the propagation range of 45 mm to 155 mm. Solid lines correspond to the material with intrinsic loss ($n = 1.68 + j0.03$), while dashed lines represent the ideal, lossless case ($n = 1.68 + j0.00$). (b) Average conversion efficiencies of the two curved beams with the variations of the spacing distance of the diffractive layers.

Figure 8(b) illustrates the impact of varying the spacing distance between the two diffractive layers on the average conversion efficiencies of the two curved beams over the propagation range from 45 mm to 155 mm. At the designed target spacing of 20 mm, both curved beams achieve their highest average conversion efficiencies, demonstrating the optimal beam shaping at this spacing. As the spacing distance deviates from this optimal value, either increasing or decreasing, the average efficiencies gradually decline. This drop indicates that misalignment in spacing alters the phase relationships between the layers, resulting in less effective energy conversion and beam shaping, which highlights the importance of precise layer positioning in z-direction to maintain peak performance.

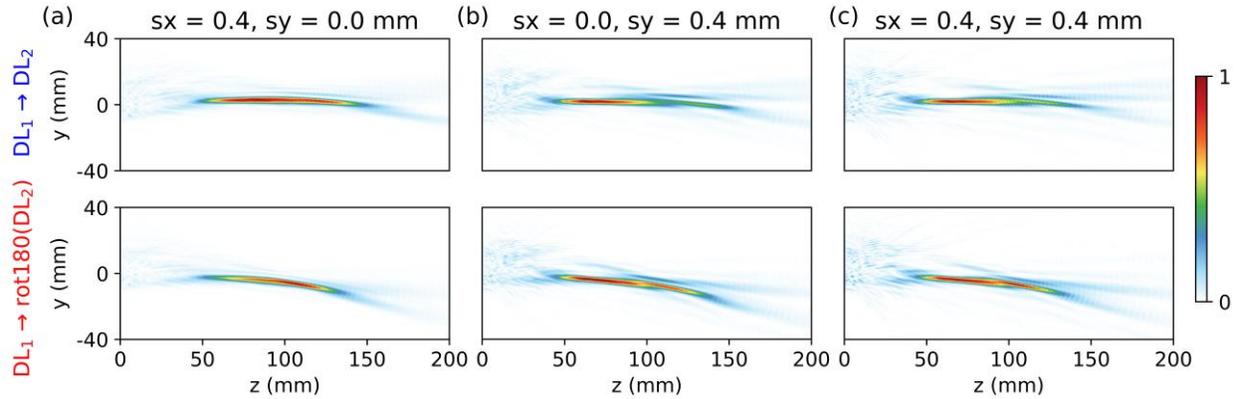

Fig. 9 Effects of spatial misalignment in x-y plane on the shape of the curved beams: (a) $DL_2$ shifted by 0.4 mm along the *x*-axis, (b) $DL_2$ shifted by 0.4 mm along the *y*-axis, (c) $DL_2$ shifted by 0.4 mm along both the *x*- and *y*-axes.

The effects of spatial misalignment on the shape of the curved beams are simulated and illustrated in Fig. 9. When the second diffractive layer ($DL_2$) is shifted by 0.4 mm along the x-axis, the curved beam profiles is largely preserved over the designed propagation range of 45-155 mm, as shown in Fig. 9(a). In contrast, when $DL_2$ is shifted by 0.4 mm along the y-axis, noticeable deformation of the curved beams occurs for propagation distances exceeding 100 mm, as shown in Fig. 9(b). As shown in Fig. 9(c), simultaneous shifts of $DL_2$ by 0.4 mm along both the x- and y-axes produce a degradation behavior similar to that observed for the y-axis shift alone, indicating that for this level of misalignment, the structure is more sensitive to misalignment along the y-direction than along the x-direction.

## 5. Conclusion

We have demonstrated a passive and reconfigurable approach for generating curved terahertz beams with predetermined propagation trajectories using inverse-designed bilayer diffractive optical elements. By spatially multiplexing two phase-only layers and exploiting mechanical rotation for reconfiguration, distinct curved beam paths are achieved without redesigning or refabricating the individual diffractive components. Unlike analytically defined self-accelerating beams such as Airy beams, the proposed method enables arbitrary trajectory engineering directly in real space, offering greater flexibility for THz applications in the radiative near-field. The experimental results at 0.3 THz validate the ability of the bilayer structure to maintain controlled beam confinement along the designed paths. This work establishes inverse-designed multilayer diffractive optics as a scalable and low-complexity platform for reconfigurable curved-beam generation, with potential impact on obstacle-aware THz wireless links, adaptive imaging, and near-field wavefront manipulation.

**Acknowledgements**

The authors acknowledge funding support from the National Science Foundation (#2234413 and #2420706).

**Conflict of Interest**

The authors declare no conflict of interest.

**Data Availability Statement**

All the data and methods needed to evaluate the conclusions in this work are present in the main text and the Supplementary Information. Any other relevant data is available from the authors upon reasonable request.

## Code Availability

The code used in this work is based on publicly available standard libraries, primarily implemented using PyTorch.